\documentclass[preprint]{elsarticle}
\usepackage{lineno}
\usepackage{graphicx}

\newcommand{\affuni}[2]{Dipartimento di Fisica dell'Universit\`a #1, #2, Italy.}
\newcommand{\affinfn}[2]{INFN Sezione di #1, #2, Italy.}
\begin{document}
\begin{frontmatter}
\title{Measurement of the branching fraction for the decay $K_S \to \pi \mu \nu$
with the KLOE detector \vspace{0.5cm} \newline \small The KLOE-2 Collaboration}
\author[Frascati]{D.~Babusci}
\author[Warsaw]{M.~Berlowski}
\author[Frascati]{C.~Bloise}
\author[Frascati]{F.~Bossi}
\author[INFNRoma3]{P.~Branchini}
\author[Roma3,INFNRoma3]{A.~Budano}
\author[Uppsala]{B.~Cao}
\author[Roma3,INFNRoma3]{F.~Ceradini}
\author[Frascati]{P.~Ciambrone}
\author[Catania,Frascati]{F.~Curciarello}
\author[Cracow]{E.~Czerwi\'nski}
\author[Roma1,INFNRoma1]{G.~D'Agostini}
\author[Frascati]{E.~Dan\`e}
\author[Roma1,INFNRoma1]{V.~De~Leo}
\author[Frascati]{E.~De~Lucia}
\author[Frascati]{A.~De~Santis}
\author[Frascati]{P.~De~Simone}
\author[Roma3,INFNRoma3]{A.~Di~Cicco}
\author[Roma1,INFNRoma1]{A.~Di~Domenico}
\author[Frascati]{D.~Domenici}
\author[Frascati]{A.~D'Uffizi}
\author[Roma2,INFNRoma2]{A.~Fantini}
\author[Frascati]{P.~Fermani}
\author[ENEAFrascati,INFNRoma1]{S.~Fiore}
\author[Cracow]{A.~Gajos}
\author[Roma1,INFNRoma1]{P.~Gauzzi}
\author[Frascati]{S.~Giovannella}
\author[INFNRoma3]{E.~Graziani}
\author[BINP,Novosibirsk]{V.~L.~Ivanov}
\author[Uppsala]{T.~Johansson}
\author[Frascati]{X.~Kang}
\author[Cracow]{D.~Kisielewska-Kami\'nska}
\author[BINP,Novosibirsk]{E.~A.~Kozyrev}
\author[Warsaw]{W.~Krzemien}
\author[Uppsala]{A.~Kupsc}
\author[BINP,Novosibirsk]{P.~A.~Lukin}
\author[Messina,INFNCatania]{G.~Mandaglio}
\author[Frascati,Marconi]{M.~Martini}
\author[Roma2,INFNRoma2]{R.~Messi}
\author[Frascati]{S.~Miscetti}
\author[Frascati]{D.~Moricciani}
\author[Cracow]{P.~Moskal}
\author[Cracow]{S.~Parzych}
\author[INFNRoma3]{A.~Passeri}
\author[Energetica,INFNRoma1]{V.~Patera}
\author[Roma1,INFNRoma1]{E.~Perez~del~Rio}
\author[Frascati]{P.~Santangelo}
\author[Calabria,INFNCalabria]{M.~Schioppa}
\author[Roma3,INFNRoma3]{A.~Selce\corref{cor1}}\ead{andrea.selce@roma3.infn.it}
\author[Cracow]{M.~Silarski}
\author[Frascati,IFIN]{F.~Sirghi}
\author[BINP,Novosibirsk]{E.~P.~Solodov}
\author[INFNRoma3]{L.~Tortora}
\author[INFNPisa]{G.~Venanzoni}
\author[Warsaw]{W.~Wi\'slicki}
\author[Uppsala]{M.~Wolke}
\address[Catania]{Dipartimento di Fisica e Astronomia "Ettore Majorana",
Universit\`a di Catania, Italy}
\address[INFNCatania]{\affinfn{Catania}{Catania}}
\address[Cracow]{Institute of Physics, Jagiellonian University, Cracow, Poland.}
\address[Frascati]{Laboratori Nazionali di Frascati dell'INFN, Frascati, Italy.}
\address[IFIN]{Horia Hulubei National Institute of Physics and Nuclear Engineering, M\v{a}gurele, Romania}
\address[Messina]{Dipartimento di Scienze Matematiche e Informatiche, Scienze Fisiche e Scienze della Terra dell'Universit\`a di Messina, Messina, Italy.}
\address[BINP]{Budker Institute of Nuclear Physics, Novosibirsk, Russia.}
\address[Novosibirsk]{Novosibirsk State University, Novosibirsk, Russia.}
\address[INFNPisa]{\affinfn{Pisa}{Pisa}}
\address[Calabria]{\affuni{della Calabria}{Rende}}
\address[INFNCalabria]{INFN Gruppo collegato di Cosenza, Rende, Italy.}
\address[Energetica]{Dipartimento di Scienze di Base ed Applicate per l'Ingegneria dell'Universit\`a ``Sapienza'', Roma, Italy.}
\address[Marconi]{Dipartimento di Scienze e Tecnologie applicate, Universit\`a ``Guglielmo Marconi", Roma, Italy.}
\address[Roma1]{\affuni{``Sapienza''}{Roma}}
\address[INFNRoma1]{\affinfn{Roma}{Roma}}
\address[Roma2]{\affuni{``Tor Vergata''}{Roma}}
\address[INFNRoma2]{\affinfn{Roma Tor Vergata}{Roma}}
\address[Roma3]{Dipartimento di Matematica e Fisica dell'Universit\`a 
``Roma Tre'', Roma, Italy.}
\address[INFNRoma3]{\affinfn{Roma Tre}{Roma}}
\address[ENEAFrascati]{ENEA, Department of Fusion and Technology for Nuclear Safety and Security, Frascati (RM), Italy}
\address[Uppsala]{Department of Physics and Astronomy, Uppsala University, Uppsala, Sweden.}
\address[Warsaw]{National Centre for Nuclear Research, Warsaw, Poland.}
\cortext[cor1]{Corresponding author}
%\linenumbers
\newpageafter{author}
\begin{abstract}
Based on a sample of 300 million $K_S$ mesons produced in $\phi \to K_L K_S$ decays recorded by the KLOE experiment
at the DA$\Phi$NE $e^+e^-$ collider we have measured the 
branching fraction for the decay $K_S \to \pi \mu \nu$. 
The $K_S$ mesons are identified by the interaction of $K_L$ mesons
in the detector. The $K_S \to \pi \mu \nu$ decays are selected by
a boosted decision tree built with kinematic variables and by a time-of-flight 
measurement. 
Signal efficiencies are evaluated with data control samples of 
$K_L \to \pi \mu \nu$ decays. 
A fit to the reconstructed muon mass distribution finds $7223 \pm 180$ signal
events. Normalising to the $K_S \to \pi^+ \pi^-$ decay events   
the result for the branching fraction is 
$\mathcal{B}(K_S \to \pi \mu \nu) = (4.56 \pm 0.11_{\rm stat} \pm 0.17_{\rm syst})\times10^{-4}$. It is the first measurement of this decay mode and the result allows an independent determination of $|V_{us}|$
and a test of the lepton-flavour universality.
%It is the first measurement of this decay mode and the result is in agreement with the expected value of $(4.69 \pm 0.06) \times 10^{-4}$ assuming lepton-flavour universality.

\end{abstract}
\end{frontmatter}
%
%\linenumbers
%
\section{Introduction} \label{INTRODUCTION}
The branching fraction for semileptonic decays of charged and neutral kaons
together with the lifetime measurements are used to determine the $|V_{us}|$ 
element of the Cabibbo--Kobayashi--Maskawa 
quark mixing matrix. The relation among the matrix elements of the first row, 
$|V_{ud}|^2 + |V_{us}|^2 + |V_{ub}|^2 = 1$, 
provides the most stringent test of the unitarity of the quark mixing matrix.
Different factors contribute to the uncertainty in
determining $|V_{us}|$ from kaon 
decays~\cite{ref:Antonelli2010,ref:Moulson2017,ref:KLOE|Vus|} 
and among the six semileptonic 
decays the contribution of the lifetime uncertainty is smallest 
for the $K_S$ meson. Nevertheless, given the lack of pure high-intensity
$K_S$ meson beams contrary to the case of $K^{\pm}$ and $K_L$ mesons, 
the $K_S \to \pi e \nu$ decay provides the least precise determination of 
$|V_{us}|$, and the branching fraction $\mathcal{B}(K_S \to \pi \mu \nu)$ 
has not yet been measured.
Measurement of this decay mode allows an independent determination of $|V_{us}|$ and to extend the 
test of lepton-flavour universality to $K_S$ semileptonic decays by comparison with the expected value of 
$(4.69 \pm 0.06) \times 10^{-4}$~\cite{ref:PDG} derived from $\mathcal{B}(K_S \to \pi e \nu)$.
%the value expected from $\mathcal{B}(K_S \to \pi e \nu$) assuming lepton-flavour universality is $(4.69 \pm 0.06 )\times10^{-4}$ ~\cite{ref:PDG}.

We present a measurement of the $K_S \to \pi \mu \nu$
branching fraction performed by the KLOE experiment at the
DA$\Phi$NE $\phi$--factory of the Frascati National Laboratory
based on an integrated luminosity of 1.6 fb$^{-1}$.
DA$\Phi$NE~\cite{ref:DAFNE} is an electron--positron collider 
running at the centre-of-mass energy of 1.02 GeV colliding $e^+$ and $e^-$
beams at an angle of $\pi$$-$0.025 rad and with a bunch-crossing
period of 2.71 ns. The $\phi$ mesons are
produced with a small transverse momentum of 13 MeV
and $K_L$--$K_S$ pairs are produced almost back-to-back with a 
cross section times the $\phi\to K_L K_S$ branching fraction of about 1 $\mu$b.
The beam energy, the energy spread, the beams
transverse momentum and the position of the interaction point
are measured using Bhabha scattering 
events~\cite{ref:Luminosity}.

The $K_S$ ($K_L$) mesons are identified (\textit{tagged}) by the observation of a $K_L$ ($K_S$) 
meson in the opposite hemisphere.
This tagging procedure allows the selection
efficiency for $K_S \to \pi \mu \nu$ to be evaluated with good 
accuracy using a sample of the abundant decay $K_L \to \pi \mu \nu$ 
tagged by the detection of $K_S \to \pi^+ \pi^-$ decays. The branching fraction
is extracted normalising the number of 
$K_S \to \pi \mu \nu$ events to the number of $K_S \to \pi^+ \pi^-$ events
recorded in the same dataset. 

\section{The KLOE detector} \label{DETECTOR}

The detector\footnote{KLOE uses a coordinate system where the $z$ axis is the bisector of the electron and positron beams, $x$ and $y$ axes define the transverse plane, the polar angle is relative to the $z$ axis.} consists of a large-volume cylindrical drift chamber, 
surrounded by a lead/scintillating fibers finely-segmented calorimeter. 
A superconducting coil around the calorimeter provides a 0.52 T axial 
magnetic field. The beam pipe at the interaction region
is spherical in shape with 10 cm radius, made of a 0.5 mm thick 
beryllium-aluminum alloy. Final-focus quadrupoles are located at $\pm$50 cm 
from the interaction region. Two small lead/scintillating-tile 
calorimeters~\cite{ref:QCAL}  are wrapped around the quadrupoles.

The drift chamber~\cite{ref:DC}, 4 m in diameter and 3.3 m long, has 
12582 drift cells arranged in 58 concentric rings with alternating stereo angles 
and is filled with a low-density gas mixture of 90\% helium--10\% isobutane. 
The chamber shell is made of carbon fiber-epoxy composite with an 
internal wall of 1.1 mm thickness at 25 cm radius. The spatial
resolution is $\sigma_{xy} =$ 0.15 mm and $\sigma_z =$ 2 mm
in the transverse and longitudinal projections, respectively. 
The momentum resolution is $\sigma_{p_{\rm T}}/p_{\rm T} = 0.4\%$, tracks vertices are reconstructed with a spatial resolution of about 3 mm.

The calorimeter~\cite{ref:EMC} is divided into a barrel and two endcaps 
and covers 98\% of the solid angle. The readout granularity is 
4.4$\times$4.4 cm$^2$, for a total of 2440 cells arranged in
five layers. Each cell is read out at both ends by photomultipliers. 
The energy deposits are obtained from signal amplitudes while 
the arrival time and the position along the fibers are obtained from  time differences between the two signals. Cells close in time and space are grouped into energy
clusters. The cluster energy $E$ is the sum of the cell energies, 
the cluster time and position are energy-weighted averages. 
Energy and time resolutions are $\sigma_E/E = 0.057/\sqrt{E\ {\rm (GeV)}}$
and $\sigma_t = 54\ {\rm ps}/\sqrt{E\ {\rm (GeV)}} \oplus 100$ ps, respectively. 
The cluster spatial resolution is 
$\sigma_{\parallel} = 1.4\ {\rm cm}/\sqrt{E\ {\rm (GeV)}}$
along the fibers and $\sigma_{\perp} = 1.3$ cm in the orthogonal direction.

The first-level trigger~\cite{ref:Trigger} uses both the calorimeter and the drift chamber 
information; the calorimeter trigger requires two energy deposits 
with $E > 50$ MeV in the barrel and $E > 150$ MeV in the endcaps;
the drift chamber trigger is based on the number and topology of 
hit drift cells. A second-level cosmic-ray veto rejects events with at least 
two energy deposits above 30 MeV in the outermost calorimeter
layer. The trigger time is determined by the first particle reaching
the calorimeter and is synchronised with the DA$\Phi$NE radio frequency signal.
The time interval between bunch crossings is smaller than the time
spread of the signals produced by the particles, thus the time of the bunch crossing originating the event, T$_0$, is determined after
event reconstruction and all the times related to that event are 
shifted accordingly. 
Data for reconstruction are selected by an on-line filter~\cite{ref:Datarec} 
to reject beam backgrounds. The filter also records the events into 
different output files for analysis according to their properties and topology (event classification), 5\% of the events are recorded without applying the filter to control the efficiency of the event classification. 

\section{Data sample and event preselection} \label{DATASAMPLE}
Processes of interest for the analysis are simulated with the GEANFI Monte Carlo (MC) program~\cite{ref:Datarec} for an integrated luminosity equal to that of the data. All $\phi$ decays are generated according to their branching fractions as well as other final states produced in $e^+e^-$ annihilation. 
%The value of the branching fraction for $K_S \to \pi \mu \nu$ decay is set to $4.692 \times 10^{-4}$. 
The operating conditions of DA$\Phi$NE during data taking as well as measurements of beam parameters are included in the MC on a run-by-run basis. Calorimeter energy deposits and drift chamber hits from beam background acquired with a random trigger are overlaid onto the simulated events. 
%Processes of interest for the analysis are simulated with the \texttt{GEANFI}~\cite{ref:Datarec} program for an integrated luminosity equal to that of the data.  Calorimeter energy deposits and drift chamber hits from beam background events triggered  at random are overlaid onto the simulated events. 
The simulated
events are processed with the same reconstruction algorithms
as the data.
 
Kaons from $\phi$-meson decays are emitted in two opposite hemispheres
with mean decay path $\lambda_S = 5.9$ mm and $\lambda_L = 3.4$ m,
thus about 50\% of $K_L$ mesons reach the calorimeter before decaying.
The velocity of the $K_L$ in the $\phi$ reference system is $\beta^* = 0.22$.
$K_S$ mesons are tagged by $K_L$ interactions 
in the calorimeter, named $K_L$-crash in the following, with a clear signature of a late signal of about 25 ns not associated to tracks.
The following requirements are applied to select $K_L$-crash:
\begin{itemize}
\item a cluster with energy $E_{\rm clu} > 100$ MeV not associated to tracks
(neutral cluster); the centroid of the neutral cluster defines the $K_L$ 
direction with a resolution of $\sim$1$^{\circ}$; 
\item polar angle of the neutral cluster 
$15^{\circ} < \theta_{\rm clu} < 165^{\circ}$ to suppress small-angle beam 
backgrounds;
\item $0.17 < \beta^* < 0.28$ for the velocity in the $\phi$ reference 
system of the particle originating the neutral cluster; $\beta^*$ is obtained from 
the velocity in the laboratory system, 
$\beta = r_{\rm clu} / c t_{\rm clu}$, with $t_{\rm clu}$ being the cluster time 
and $r_{\rm clu}$ the distance from the nominal interaction point,
the $\phi$ transverse momentum determined run-by-run and the angle between the $\phi$ momentum and 
the $K_L$-crash direction.
\end{itemize}
%These requirements are used to tag $K_S$ mesons. 
Assuming the neutral kaon mass, the $K_S$ 4-momentum is defined by the 
$K_L$-crash direction and the $\phi$ 4-momentum:
$P_{K_S} = P_{\phi} - P_{K_L}$. 

The $K_S \to \pi \mu \nu$ candidates are selected
requiring two tracks of opposite curvature forming a vertex inside the cylinder
defined by 
\begin{equation}
\rho_{\rm vtx} = \sqrt{x^2_{\rm vtx}  + y^2_{\rm vtx}} < 5\ {\rm cm}  
\qquad |z_{\rm vtx}| < 10\ {\rm cm} .
\label{eq:Vertex}
\end{equation}
In case more than one vertex is found, the closest to the interaction region is chosen. The above requirements define the event preselection.

After preselection, the data sample 
contains about 300 million events and its composition, 
as evaluated by simulation, is shown in Table~\ref{tab:01}.
The large majority of events are $K_S \to \pi^+ \pi^-$ decays, and there is also a
large contribution from $\phi \to K^+ K^-$ events where one  kaon
or its decay products generate a fake $K_L$-crash and the other kaon 
decays early into $\pi^{\pm}\pi^0$.
%=========Table 1
\begin{table}[htp]
\caption{Number of data and simulated events after preselection.}
\begin{center}
\begin{tabular}{lrr}
& Events [$10^3$] & Fraction [\%] \\ 
\hline
Data & 301646 \\
MC   & 312019 \\
$K_S \to \pi^+\pi^-$ & 301976 &  96.78\\
$\phi \to K^+K^-$ & 9566 & 3.07 \\
$K_S \to \pi e \nu$ & 259 & 0.08 \\
$K_S \to \pi \mu \nu$ &140 &    0.04    \\
$K_S \to \pi^0\pi^0$ & 30 &  0.01   \\
Others & 47 & 0.02 \\
\hline
\end{tabular}
\end{center}
\label{tab:01}
\end{table}
 
\begin{figure}[htb!]
 \centering
 \includegraphics[width = 8cm]{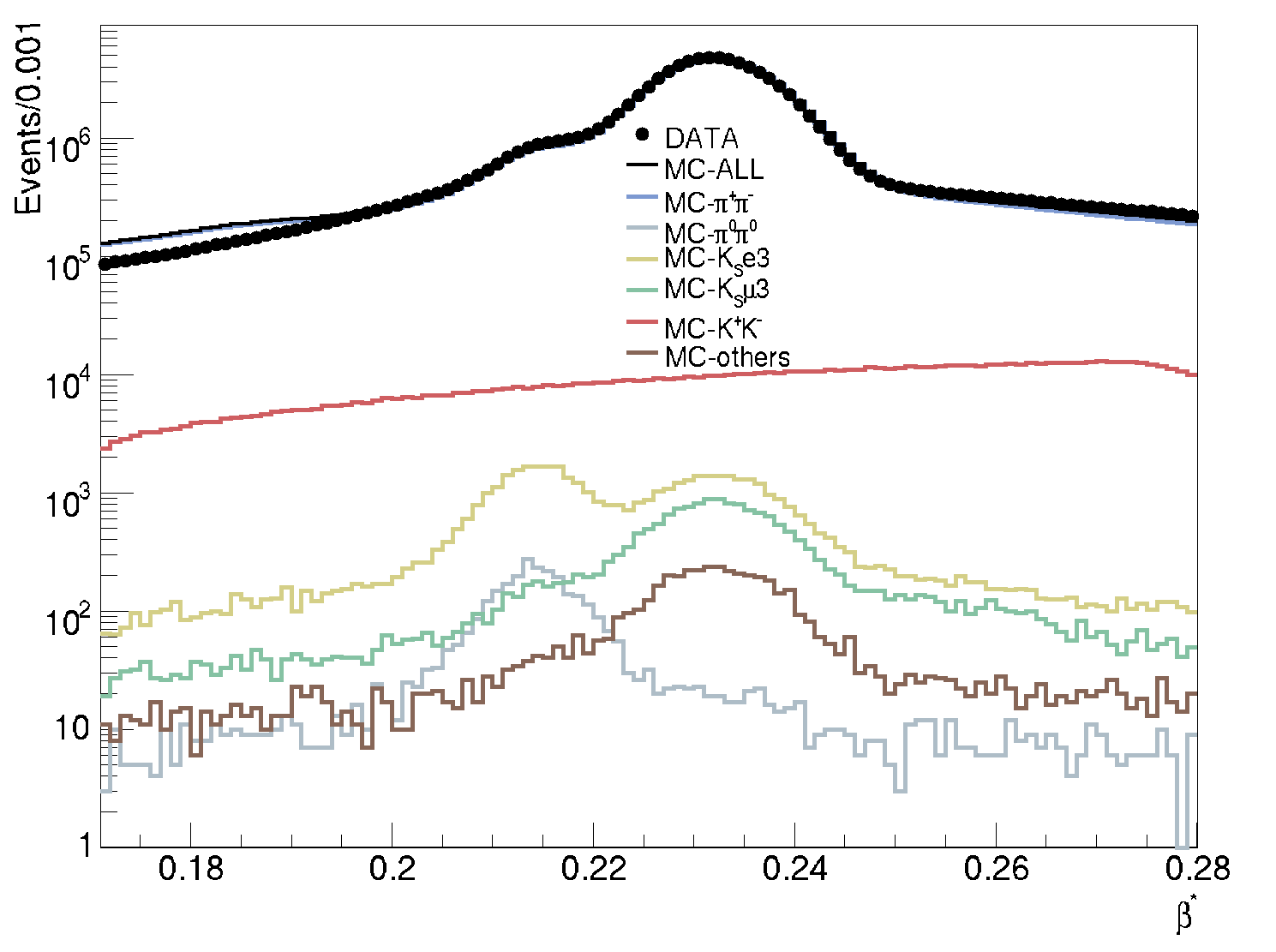}
 \caption{Distribution of $\beta^*$ after preselection for data and simulated  events.}
 \label{fig:Beta}
\end{figure}

The distribution of $\beta^*$ is shown in Figure~\ref{fig:Beta}
for data and simulated events. 
Two peaks are visible, the first is associated to events triggered by 
photons or electrons, and the second to events triggered
by charged pions. The trigger is synchronised with the bunch crossing and
the time difference between a photon (or electron) and a pion (or muon)
arriving at the calorimeter corresponds to a time shift of about one bunch-crossing.

%=========Figure 1

\section{Selection of signal and normalisation events} \label{SELECTION}

The selection of signal events is performed in two steps; first a selection
based on the event kinematics using only tracking variables and then a selection
based on the time-of-flight measured with the calorimeter. The two groups of variables
are uncorrelated. In order to assign a time to the particles 
each track is associated to a cluster. The track-to-cluster association 
(TCA) is applied as follows: for each track connected to the vertex a cluster with 
$E_{\rm clu} > 20$ MeV and $15^{\circ} < \theta_{\rm clu} < 165^{\circ}$ 
is required whose centroid is within 60 cm of the track extrapolation to the 
calorimeter front surface. The event is retained only if TCA is satisfied by both tracks. 

\begin{figure}[htb!]
 \centering
 \begin{tabular}{@{}cc@{}}
 \includegraphics[width = 5.cm]{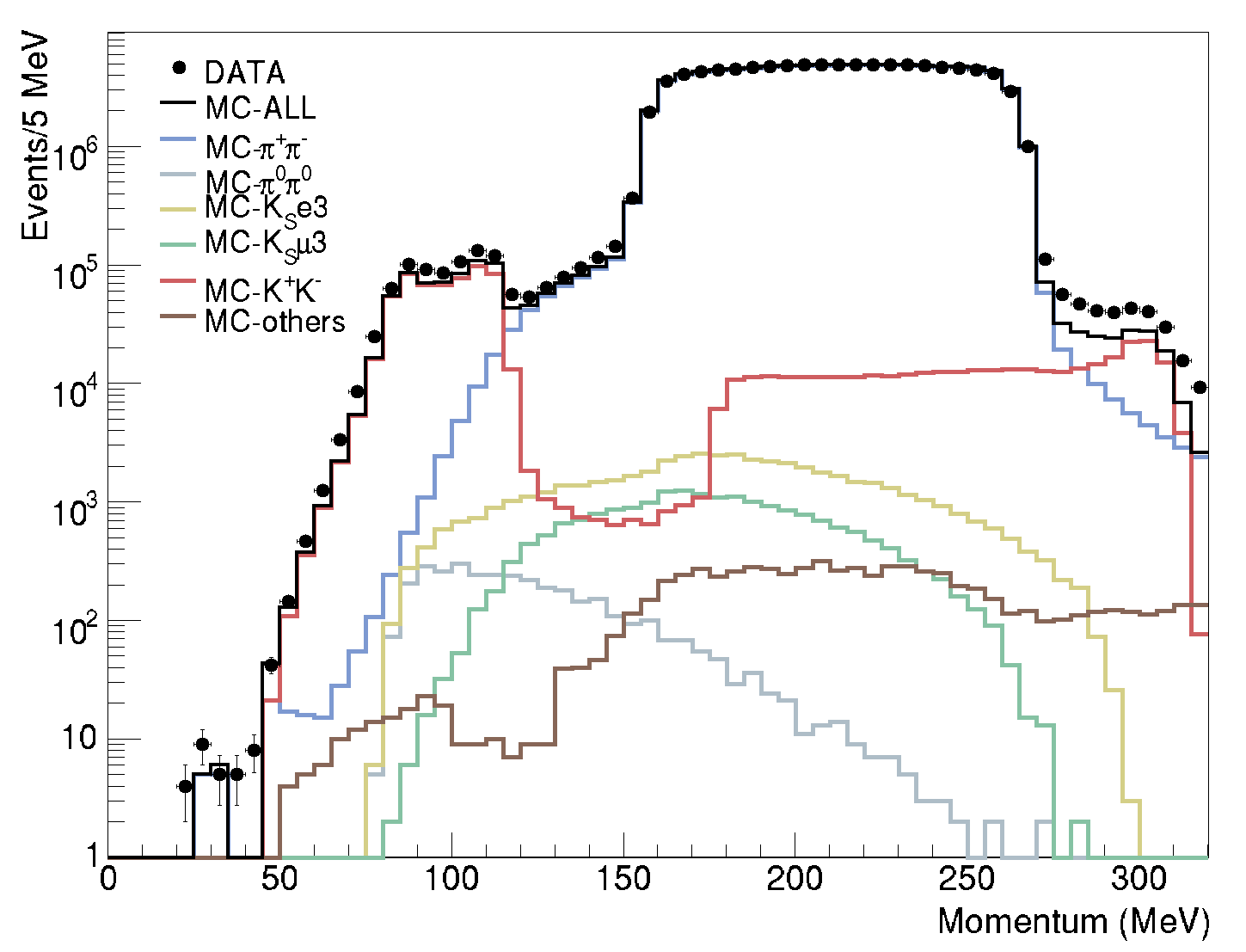}&
 \includegraphics[width = 5.cm]{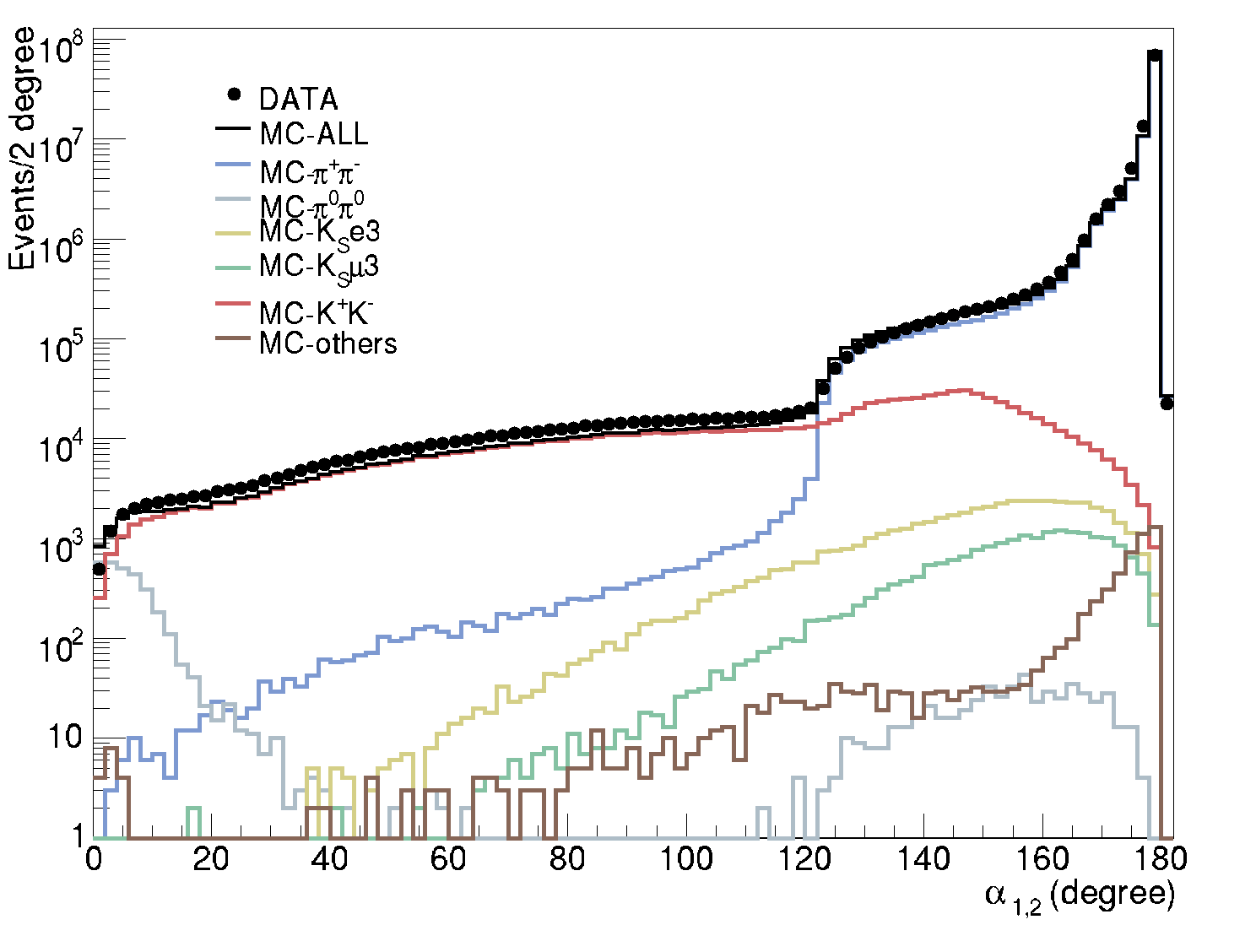}\\
 \includegraphics[width = 5.cm]{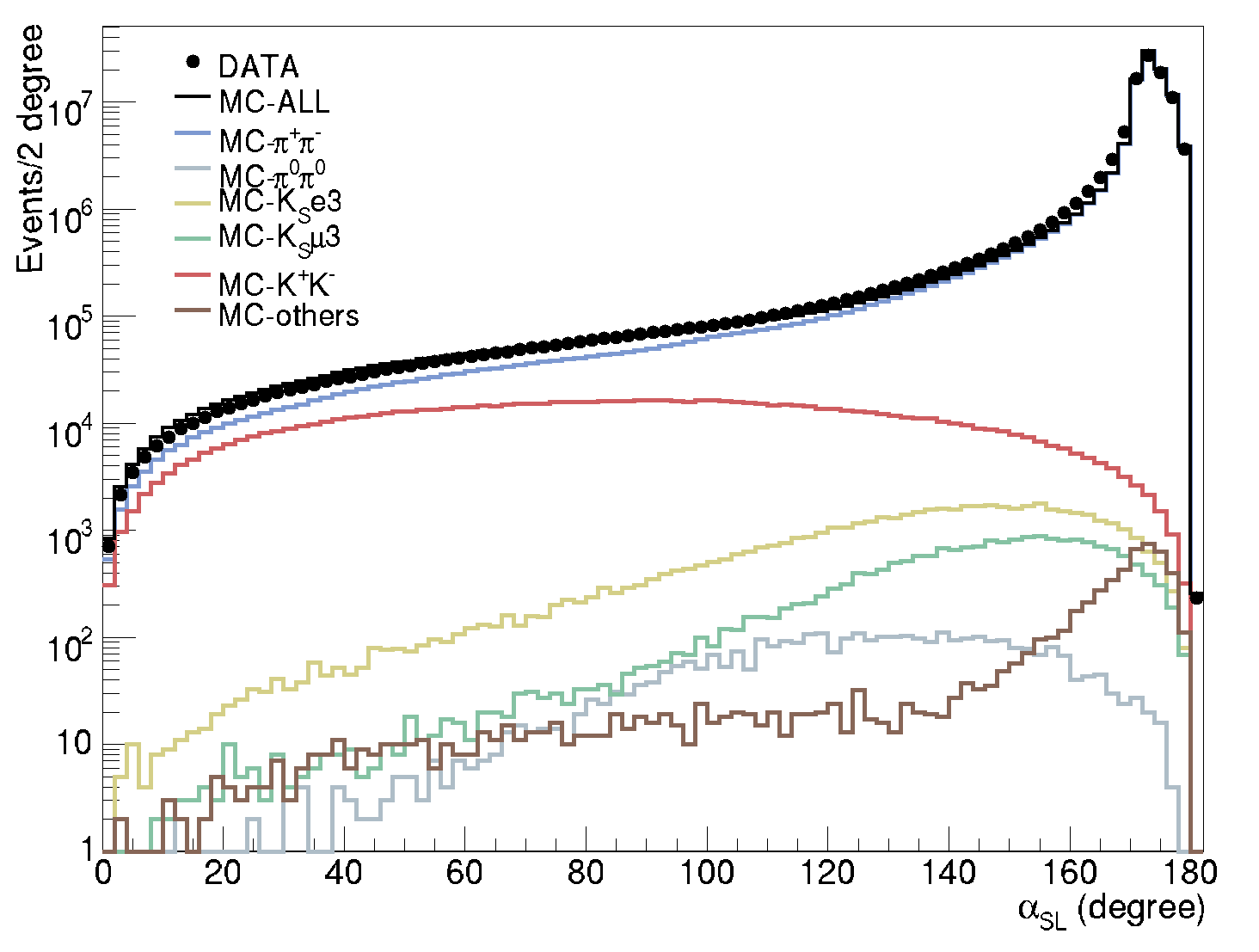}&
 \includegraphics[width = 5.cm]{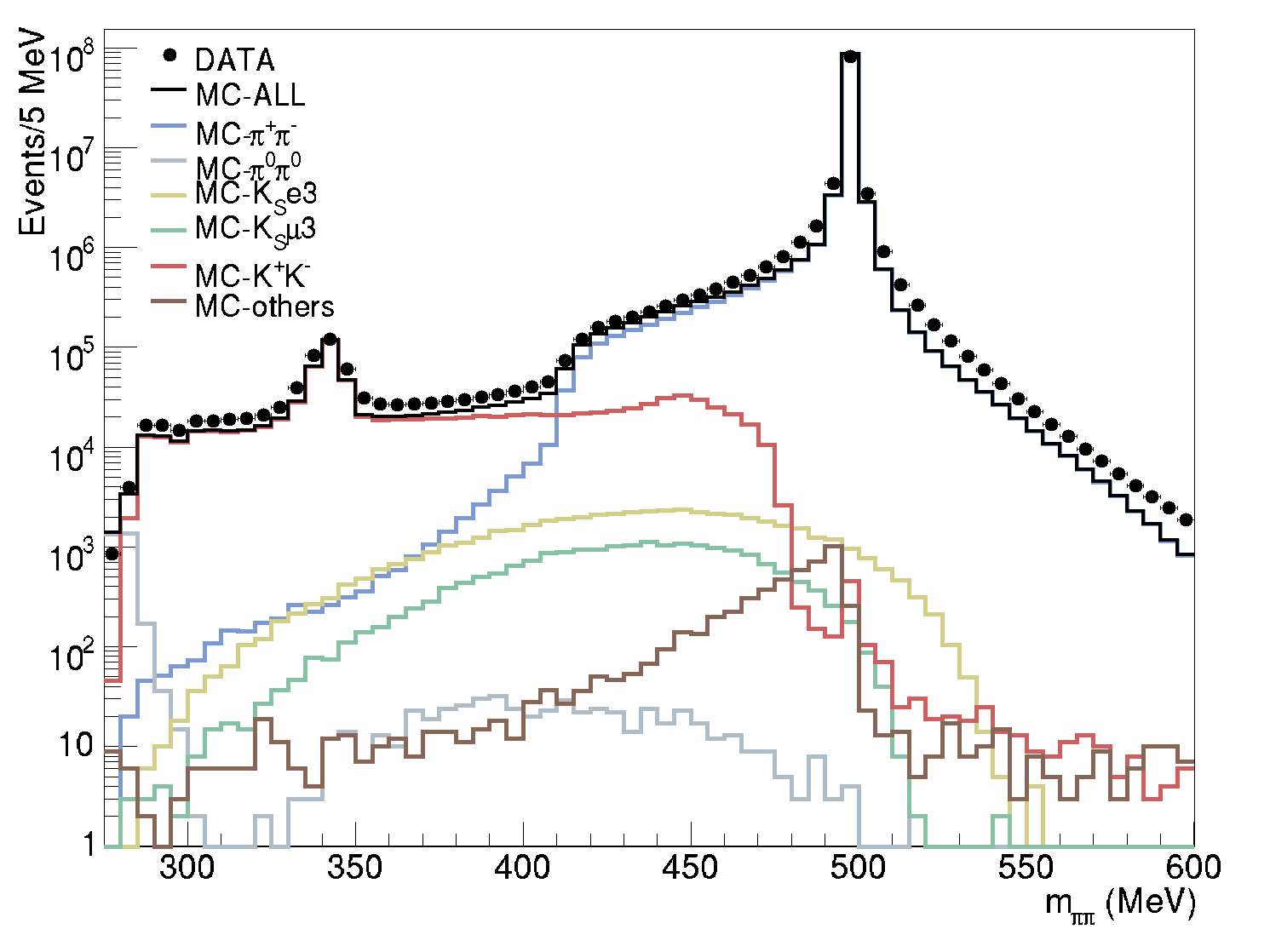}\\
 \includegraphics[width = 5.cm]{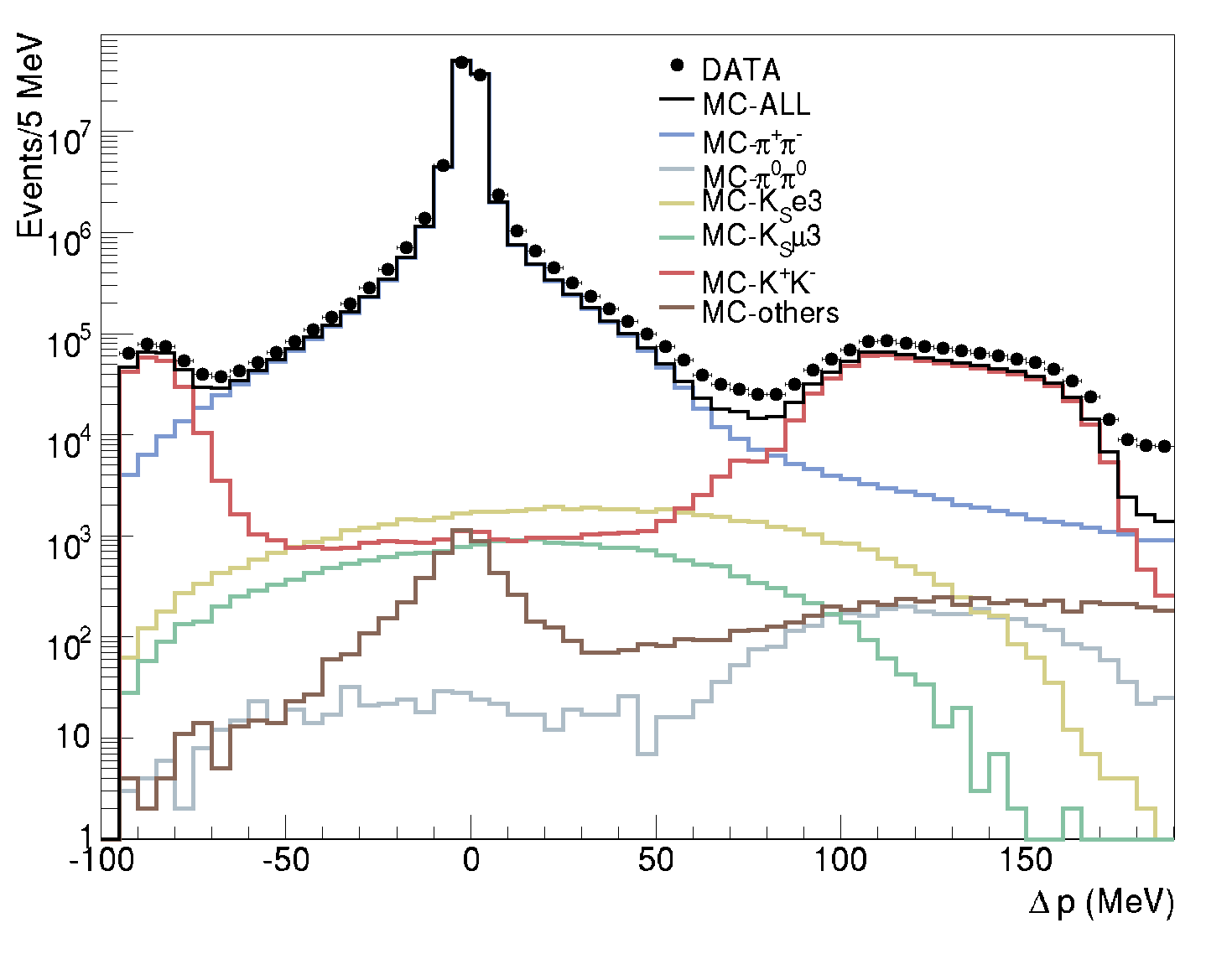}\\
  \end{tabular}
 \caption{Distributions of the variables used in the multivariate analysis for 
 data and simulated events after preselection.
 From top left: 
 track momenta ($p_1, p_2$),
angle between the two tracks in the $K_S$ reference system ($\alpha_{1,2}$), 
angle beween $K_L$ and $K_S$ directions ($\alpha_{SL}$), 
two-track invariant mass in the hypothesis of charged pions ($m_{\pi\pi}$), 
$\Delta p = |\vec{p}_{\rm sum}| - |\vec{p}_{K_S}|$.}  
 \label{fig:Variables}
\end{figure}

Five variables with good discriminating power against background are used
in a multivariate analysis. A boosted decision tree (BDT) classifier is built with the 
following variables:
\begin{itemize}
\item[$\vec{p}_1 , \vec{p}_2$] : the tracks momenta;
\item[$\alpha_{1,2}$] : the angle at the vertex between the two momenta 
in the $K_S$ reference system;
\item[$\alpha_{SL}$] : the angle between 
$\vec{p}_{\rm sum} = \vec{p}_1 + \vec{p}_2$ and the $K_L$-crash direction;
\item[$\Delta p$] : the difference between $|\vec{p}_{\rm sum}|$ and the absolute value $|\vec{p}_{K_S}|$ of the $K_S$ 
momentum determined using the tagging $K_L$;
\item[$m_{\pi\pi}$] : the invariant mass reconstructed 
from $\vec{p}_1$ and $\vec{p}_2$, in the hypothesis of charged-pion mass. 
\end{itemize}
The distributions of the variables are shown in Figure~\ref{fig:Variables} 
for data and simulated events.

%=========Figure 2
After preselection two cuts are applied to suppress the background in the tails of the distributions:
\begin{equation}
p < 320 {\rm \ MeV \ for \ both \ tracks}
\quad {\rm and} \quad
\Delta p < 190 {\rm \ MeV .}
\label{eq:ADDITIONALcut}
\end{equation}

\begin{figure}[htb!]
 \centering
 \includegraphics[width = 7.cm]{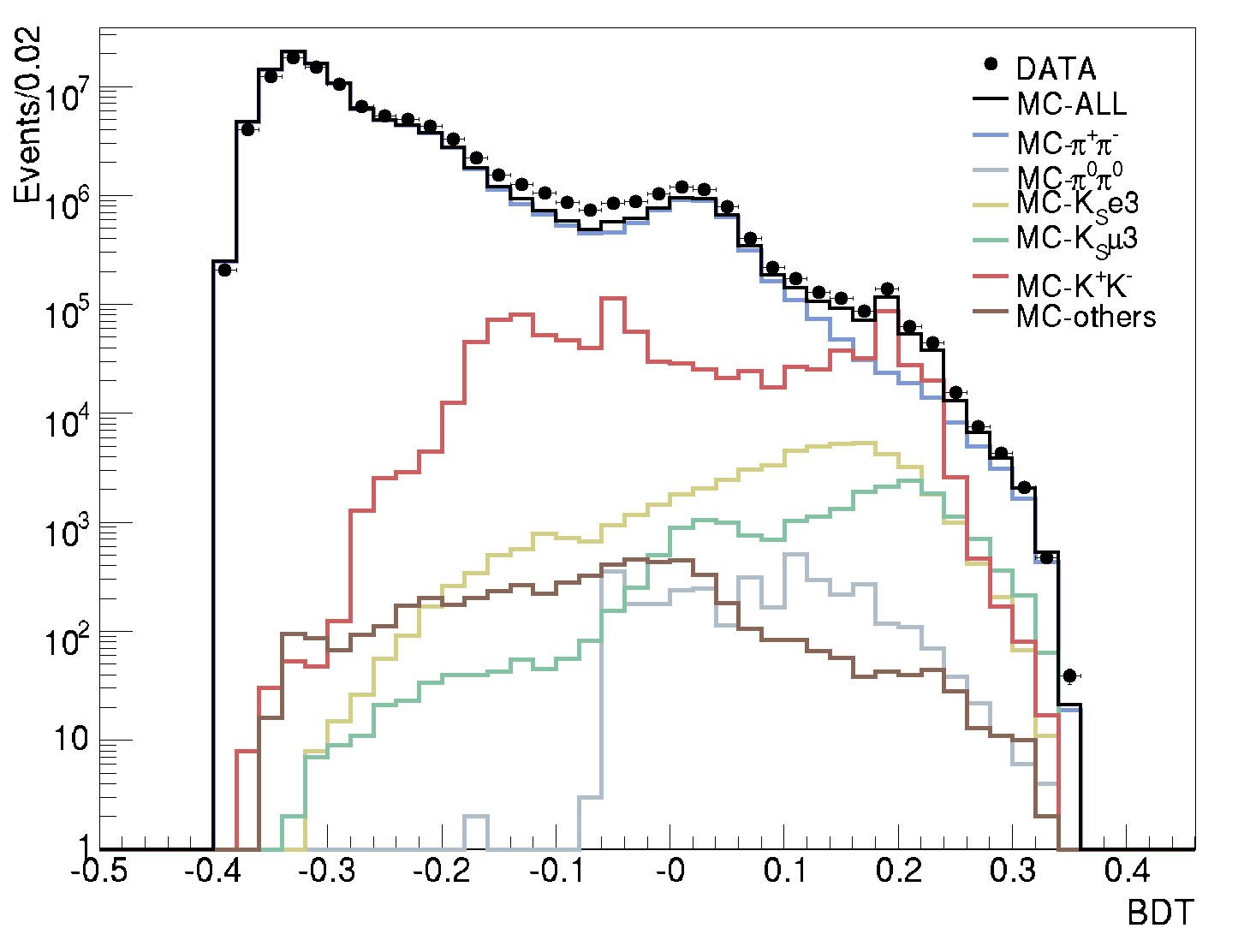}
 \caption{Distribution of the BDT classifier output for data and simulated events.}
 \label{fig:BDToutput}
\end{figure}

The training of the BDT classifier is done on a simulated sample of 5,000 
$K_S \to \pi \mu \nu$ events and a sample of 50,000 background events; 
samples of the same size are used for the test. 
After training and test the classification is run on all events of the 
MC and data sample. The distribution of the BDT classifier output is shown in 
Figure~\ref{fig:BDToutput} for data and simulated events. 
The data distribution is well reproduced by simulation in the region populated
by the signal. To suppress the large background of $K_S \to \pi^+ \pi^-$ and 
$\phi \to K^+ K^-$ events, a cut is applied
\begin{equation}
BDT > 0.18  
\label{eq:BDTcut}
\end{equation}
chosen to maximise the ratio $S/\sqrt{S+B}$ where $S$ and $B$ are the signal and background yields.

%=========Figure 3

\begin{figure}[htb!]
\begin{center}
 \centering
% \begin{tabular}{@{}cc@{}}
 \includegraphics[width = 6.cm]{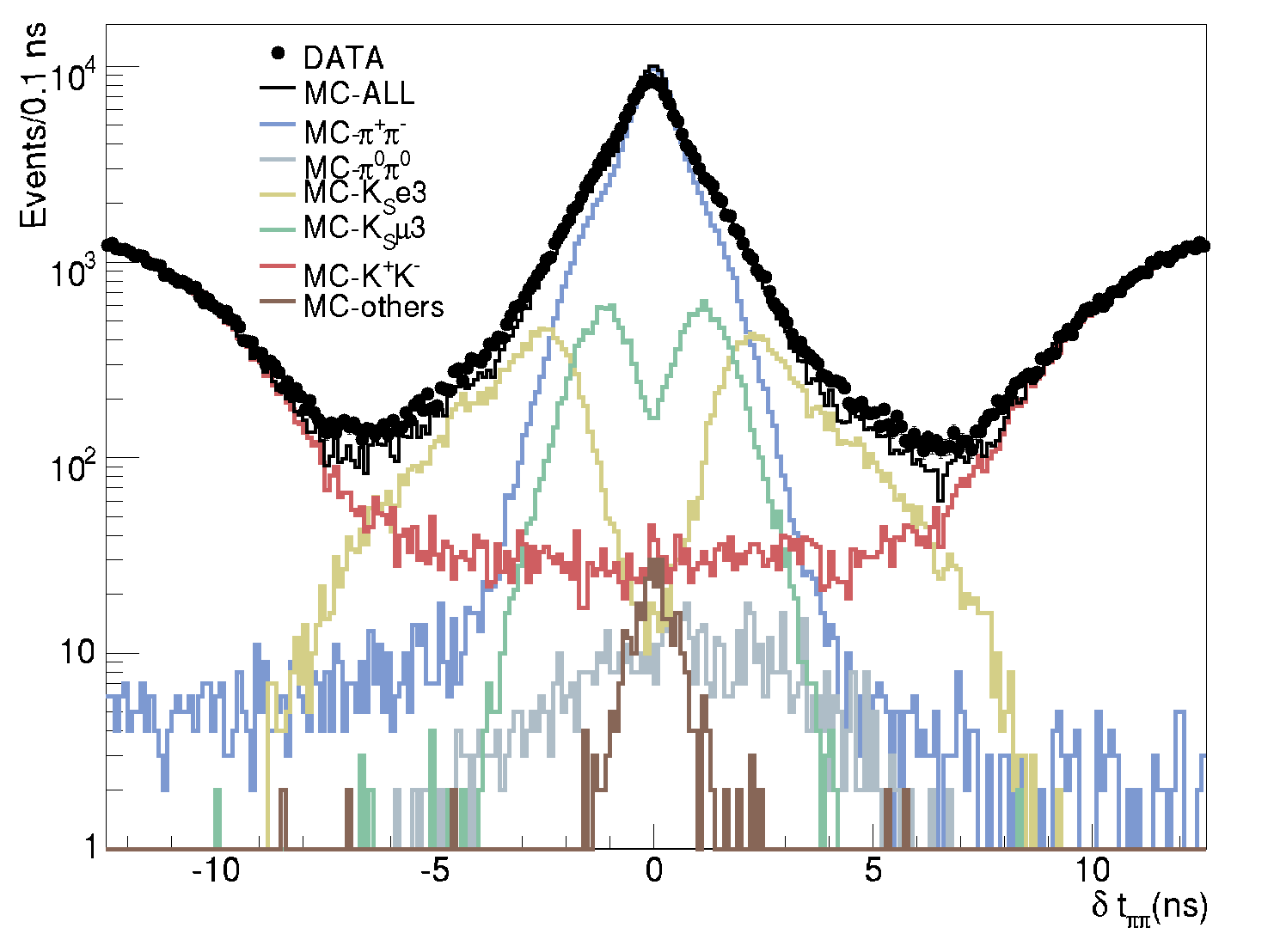}
 \includegraphics[width = 6.cm]{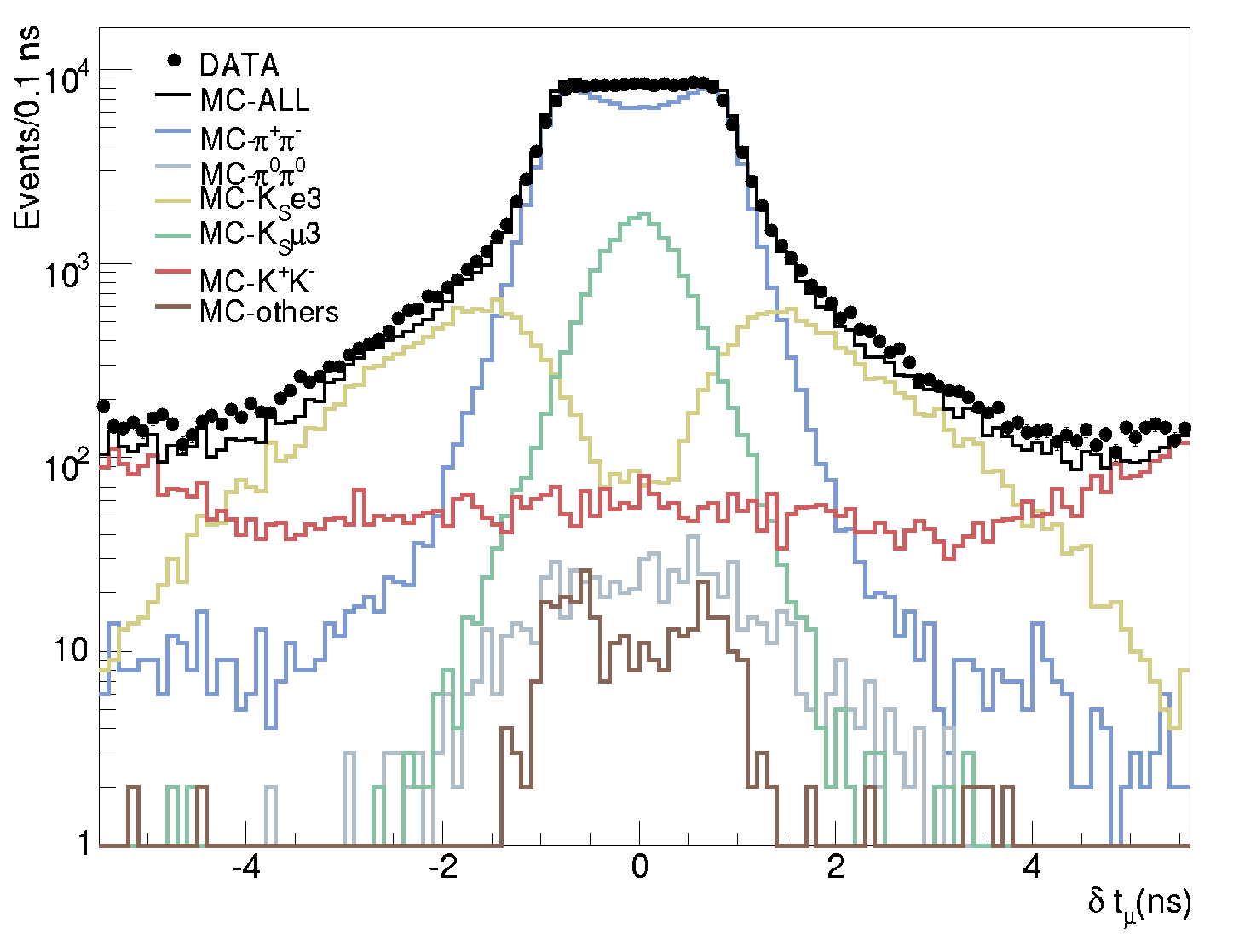}
%  \end{tabular}
 \caption{Distributions of $\delta t_{\pi\pi}$ (left) and 
 $\delta t_{\mu}$ (right)
 for data and simulated events.}
 \label{fig:dTOFpipi}
\end{center}
\end{figure}

 The selected events contain $\pi\pi$, $K\pi$, $e\pi$ track pairs for the main backgrounds and $\mu\pi$ for the signal. A selection
based on time-of-flight measurement is performed to identify $\mu\pi$ pairs.
For each track associated to a cluster, the difference between the time measured by the calorimeter and the time-of-flight measured 
along the particle trajectory 
\[ \delta t_i = t_{{\rm clu},i} - L_i / c \beta_i \qquad i = 1, 2 \]
is computed, where $t_{{\rm clu},i}$ is the time of the cluster associated to track $i$, $ L_i$ is the 
length of the track, and $\beta_i = p_i/\sqrt{p_i^2 + m_i^2}$ is function 
of the mass hypothesis for track $i$. 
%The times $t_{{\rm clu},i}$ are referred to the trigger and the same T$_0$ value is assigned to both clusters. 
To reduce the uncertainty due to the T$_0$ determination, the difference 
\[ \delta t_{1,2} = \delta t_1 - \delta t_2  \]
is used to determine the mass assignment to the tracks. 
The $\pi\pi$ hypothesis is tested first, the distribution of 
$\delta t_{\pi\pi} = \delta t_{1,\pi} - \delta t_{2, \pi}$ 
is shown in Figure~\ref{fig:dTOFpipi}(left). A fair agreement is observed 
between data and simulation,
the $K_S \to \pi \mu \nu$ and $K_S \to \pi e \nu$ distributions are well separated
and the $K^+ K^-$ background is isolated in the tails of the distribution,
however the signal is hidden under a large $K_S \to \pi^+ \pi^-$ background. 
To reduce the background a cut is applied
\begin{equation}
1\ {\rm ns} < |\delta t_{\pi\pi}| < 3\ {\rm ns} .
\label{eq:deltaTOFpipicut}
\end{equation}

%=========Figure 4
The $\pi\mu$ hypothesis 
is tested by assuming the pion and muon mass for either track
\[ \delta t_{\pi\mu} = \delta t_{1,\pi} - \delta t_{2,\mu}
\qquad {\rm and} \qquad
\delta t_{\mu\pi} = \delta t_{1,\mu} - \delta t_{2,\pi} . \] 
The two-dimensional $(\delta t_{\pi\mu},\delta t_{\mu\pi})$
distribution for simulated signal events shows that the correct mass assignment corresponds to the smaller absolute value
of the two hypotheses.
The distribution of the signed value of
 $\delta t_{\mu} = {\rm min}\left[ |\delta t_{\pi\mu}| , |\delta t_{\mu\pi}| \right]$ 
is shown in Figure~\ref{fig:dTOFpipi}(right)
for data and simulated events.  
The distribution for the signal is narrow and peaked at zero while it is broader for the
backgrounds. 
A final cut is applied
\begin{equation}
| \delta t_{\mu} | < 0.5\ {\rm ns} .
\label{eq:deltaTOFmu}
\end{equation}

%=========Table 2
\begin{table}[htp]
\caption{Number of events after the $\delta t_{\mu}$ selection for data 
and simulated events.}
\begin{center}
\begin{tabular}{lrc}
& Events & Fraction [ \% ] \\					
\hline
Data & 38686 & \\
MC   & 36444 & \\
$K_S \to \pi^+\pi^-$   & 25853 & 70.9   \\
$\phi \to K^+K^-$       & 475       & \ \ 1.30     \\
$K_S \to \pi e \nu$     & 448       & \ \ 1.23      \\
$K_S \to \pi \mu \nu$ & 9424   & 25.9      \\
Others                        &   244     &  \ \ 0.7   \\
\hline
\end{tabular}
\end{center}
\label{tab:mu003}
\end{table}
The number of surviving events in the data sample is 38686 and its composition
as evaluated by simulation is listed in Table~\ref{tab:mu003}.
After the mass assignment to the two tracks the invariant mass of the charged particle identified as the muon is evaluated as
\[ m_{\mu}^2 = \left( E_{K_S} - E_{\pi} - p_{\rm miss} \right)^2 - p^2_{\mu} \]
with $p_{\rm miss}^2 = (\vec{p}_{K_S} - \vec{p}_{\pi} - \vec{p}_{\mu})^2$,
$E_{K_S}$ and $\vec{p}_{K_S}$ being the energy  
and momentum reconstructed using the tagging $K_L$, and $\vec{p}_{\pi}$, 
$\vec{p}_{\mu}$, the momenta of the candidate pion and muon track.

\begin{figure}[htb!]
 \centering
 %\begin{tabular}{@{}cc@{}}
 \includegraphics[width = 6.cm]{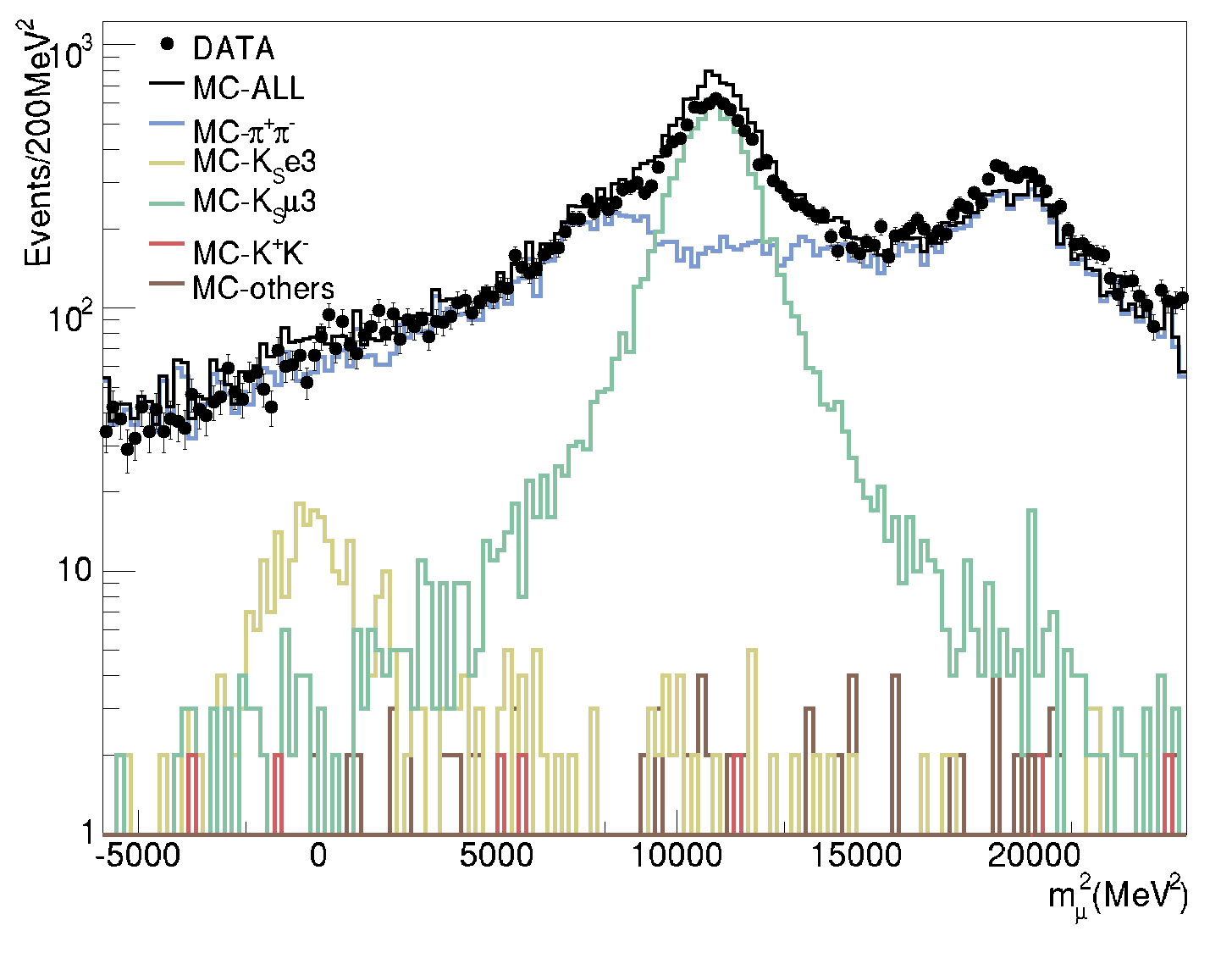}
 \includegraphics[width = 6.cm]{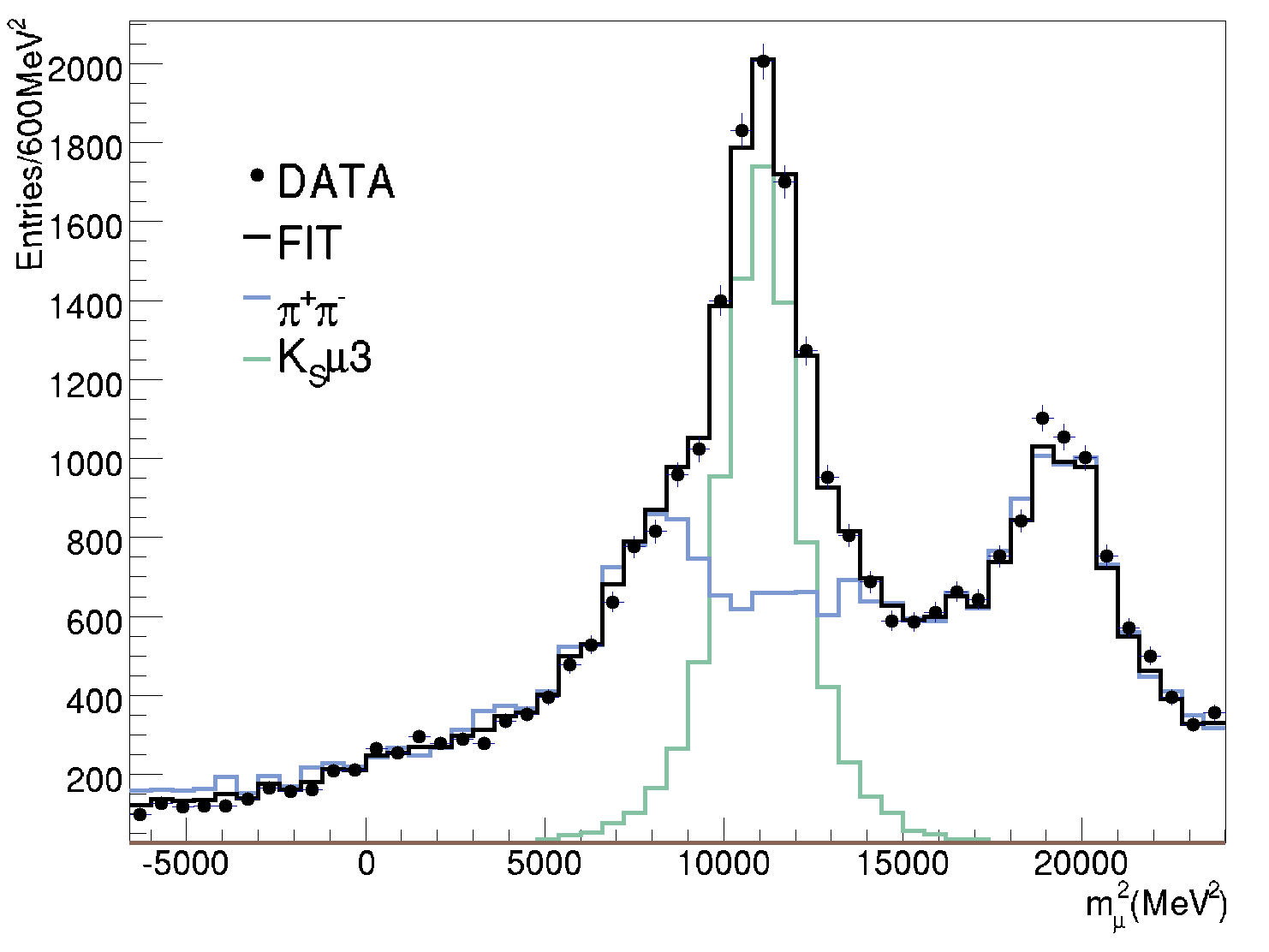}
%  \end{tabular}
 \caption{The $m^2_{\mu}$ distribution for data, MC signal and background (left); 
 comparison of data with the fit (right).}
 \label{fig:m2FIT}
\end{figure}

The number of signal events is extracted with a fit to the 
$m_{\mu}^2$ distribution with the MC shapes of three components:
$K_S \to \pi \mu \nu$, $K_S \to \pi^+\pi^-$ and the sum of all other backgrounds. 
The fit is performed in the range $-6000 < m^2_{\mu} < 24000$ MeV$^2$
with 48 degrees of freedom. The third component, which
is peaked around $m_e^2$, is constrained to a negligible value by the fit. 
Figure~\ref{fig:m2FIT} shows the distribution of $m_{\mu}^2$ 
for data, simulated events and the fit, and Table~\ref{tab:FitOutput} 
presents the result of the fit. The number of signal events is
\[ N_{\pi\mu\nu} = 7223 \pm 180 \qquad {\rm with} \ \chi^2/{\rm ndf} = 30/48 . \]

%=========Figure 5

%=========Table 3
\begin{table}[htp]
\caption{Result of the fit to the $m^2_\mu$ distribution.}
\begin{center}
\begin{tabular}{lccc}
\hline
& Fraction & Events \\ 
\hline
$\pi \mu \nu $ & 0.23  & \ 7223 $\pm$ 180  \\
$\pi^+ \pi^- $   & 0.77  &  23764 $\pm$ 270  \\
\hline
Total &   & 30987 \ \qquad \qquad  \\
\hline
\end{tabular}
\end{center}
\label{tab:FitOutput}
\end{table}

The normalisation sample of 
$K_S \to \pi^+ \pi^-$ events is selected by requiring $140 < p < 280$ MeV
for both tracks (Figure~\ref{fig:Variables}). 
This requirement selects $N_{\pi \pi} = (282.314 \pm 0.017)\times10^6$ events
with a purity of 99.9\% as determined by simulation. 

\section{Determination of efficiencies} \label{EFFICIENCY}

The branching fraction for the $K_S \to \pi \mu \nu$ decay is 
evaluated as
\begin{equation}
\mathcal{B}(K_S \to \pi \mu \nu) = 
\frac{N_{\pi \mu \nu}}{\epsilon_{\pi \mu \nu}} \times
\frac{\epsilon_{\pi \pi}}{N_{\pi \pi}} \times R_{\epsilon} \times
\mathcal{B}(K_S \to \pi^+ \pi^-) ,
\label{eq:BRATIO}
\end{equation}
where $N_{\pi \mu \nu}$ and $N_{\pi \pi}$ are the numbers of 
$K_S \to \pi \mu \nu$ and $K_S \to \pi^+ \pi^-$ events, 
$\epsilon_{\pi \mu \nu}$ and $\epsilon_{\pi \pi}$
are the respective selection efficiencies, and 
$R_{\epsilon}$ is the
ratio of the efficiencies for the trigger, on-line filter and preselection for the two decays.

The signal selection efficiency is determined with $K_L \to \pi \mu \nu$ control samples (CS) and evaluated as
\begin{equation}
\epsilon_{\rm \pi \mu \nu}=\epsilon_{\rm CS} \times 
\frac {\epsilon^{\rm MC}_{\rm \pi \mu \nu}}{\epsilon^{\rm MC}_{\rm CS}} ,
\label{eq:EFFICIENCY}
\end{equation}
where $\epsilon_{\rm CS}$ is the efficiency of the control 
sample and $\epsilon^{\rm MC}_{\pi \mu \nu}$,
$\epsilon^{\rm MC}_{\rm CS}$ are the efficiencies obtained from 
simulation for the signal and control samples, respectively. 

The $K_L \to \pi \mu \nu$ decay~\cite{ref:KLtoany,ref:KLtopimunu}  
is kinematically identical to the signal, the only difference being the much 
longer decay path.
For the control sample the tagging is done with $K_S \to \pi^+ \pi^-$ decays, 
preselected in the same way as for the signal sample with the addional cut 
$|m_{\pi\pi} - m_{K^0}| < 15$ MeV to increase the purity.
The radial distance of the $K_L$ vertex is 
required to be smaller than 5 cm to match the signal selection,
but greater than 1 cm to minimise the ambiguity in identifying the 
$K_L$ and $K_S$ vertices.
The control sample is composed mainly of $K_L \to \pi e \nu$ , 
$K_L \to \pi^+\pi^-\pi^0$  and $K_L \to \pi \mu \nu$ 
 decays, while most of $K_L \to \pi^0\pi^0 \pi^0$ 
decays are rejected by the requirement of two tracks.
The distribution of the missing mass, $m^2_{\rm miss}$, of the two tracks 
connected to the $K_L$ vertex, assigning the charged-pion mass, shows a narrow isolated
peak at the $\pi^0$ mass; a cut 
$m^2_{\rm miss} < 15000$ MeV$^2$ efficiently rejects the 
$K_L \to \pi^+\pi^-\pi^0$ decays. The number of events in the control sample is 911757.

In order to evaluate the signal selection efficiency, 
two control samples are used, one selected based on kinematic variables and the other based on 
time-of-flight (TOF), the two groups of variables being largely uncorrelated. 
%The selections of the two control samples are made as follows:
%\begin{itemize}
%\item a cut on the TOF variables is applied to evaluate the %efficiency of the 
%selection based on the kinematic variables and the BDT classifier;
%\item a cut on kinematic variables is applied to evaluate the 
%efficiency of the TCA and TOF selections.
%\end{itemize}

\begin{figure}[htb!]
 \centering
 %\begin{tabular}{@{}c@{}}
 \includegraphics[width = 12.5cm]{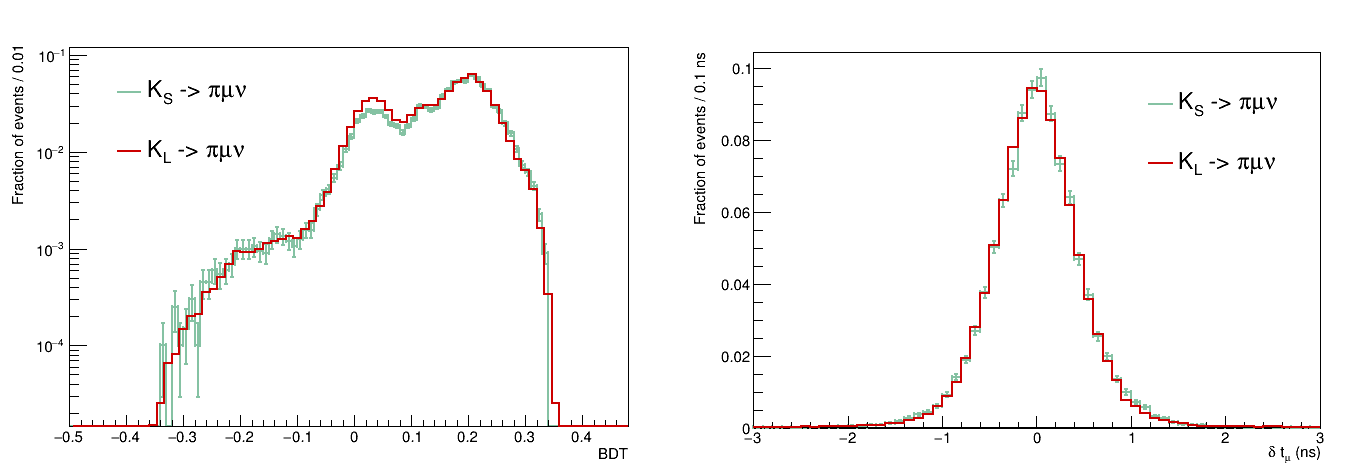}
 %\includegraphics[width = 7.7cm]{CS_dtofMU_last.png}\\
%  \end{tabular}
 \caption{Normalised Monte Carlo distributions of the BDT classifier output 
 (left) and $\delta t_{\mu}$ (right) for 
 $K_L \to \pi\mu\nu$ and $K_S \to \pi\mu\nu$ events.}
 \label{fig:BDTLS-MC}
\end{figure}

The control sample for evaluating the efficiencies of the selection with kinematic variables and BDT classifier is selected applying a cut on the two-dimensional $(\delta t_{\pi\mu},\delta t_{\mu\pi})$ distribution that removes most of the $K_L \to \pi e \nu$ events.
The purity of the sample as determined with simulation is 86\%. 
The resolutions in the measurement of the
tagging $K_S$ (control sample) are similar to those of 
the tagging $K_L$ (signal sample) and the same BDT classifier is used for both samples. 
The BDT MC distributions for the signal and control sample are compared 
in Figure~\ref{fig:BDTLS-MC}(left). 
Applying to the control sample the same selections as for the signal,
%Eqs.~(\ref{eq:ADDITIONALcut}) and~(\ref{eq:BDTcut}),
the efficiencies evaluated with Eq.~(\ref{eq:EFFICIENCY}) are
\[ \epsilon({\rm kinem.\ sel.}) = 0.982 \pm 0.004_{\rm stat} \quad {\rm and} 
\quad \epsilon({\rm BDT}) = 0.417 \pm 0.003_{\rm stat} . \]

%=========Figure 6

To evaluate  TCA and TOF efficiencies for the signal, the T0 determination using the earlier among the two clusters associated with the $K_S$ decay has to be considered. The control sample selection therefore requires the earliest cluster to be associated with one of charged secondary particles from $K_L$ decay and a cut on the $(m_{\pi\pi},m^2_{\rm miss})$  distribution to reject $K_L \to \pi e \nu$ events.
%In the $K_S \to \pi\mu\nu$ analysis  the T$_0$ is determined by the first  cluster in time, associated with one of the daughter particles  of the $K_S$ decay; then for the control sample it is required that the first cluster in time be associated with the $K_L$ decay in order not to bias the TOF variables. 
The purity of the sample as determined with simulation is 87\%. 
The MC distributions of $\delta t_{\mu}$ for the signal and control sample 
are compared in Figure~\ref{fig:BDTLS-MC}(right). 
Applying to the control sample the analysis procedure as 
for the signal the efficiencies evaluated with Eq.~(\ref{eq:EFFICIENCY}) 
are
\[ \epsilon({\rm TCA}) = 0.347 \pm 0.002_{\rm stat} \quad {\rm and} \quad
\epsilon({\rm TOF}) = 0.392 \pm 0.003_{\rm stat} . \]
The correction factors in Eq.~(\ref{eq:EFFICIENCY}) differ from one by less than 10\% but for $\epsilon$(TOF) where it differs by 20\%.

The tails of the $m^2_{\mu}$ distribution in Figure~\ref{fig:m2FIT}(left) 
are not included 
in the fit to improve its stability, the relative efficiency is $0.991 \pm 0.001$.

The signal selection efficiencies are summarised in Table~\ref{tab:EffSum}
where only the statistical errors are shown. Combining the values accounting for the correlation of the control samples we obtain $\epsilon_{\pi\mu\nu}=0.0552\pm0.0005$.

\begin{table}[htp]
\caption{Efficiencies for the signal selections. The errors are statistical, the error of the total efficiency accounts for the correlation of the control samples.}
\begin{center}
\begin{tabular}{lc}
Selection & Efficiency \\ 
\hline
Kinematic selection & 0.982 $\pm$ 0.004 \\
BDT selection & 0.417 $\pm$ 0.003 \\
TCA selection & 0.347 $\pm$  0.002 \\
TOF selection & 0.392 $\pm$ 0.003 \\
FIT range & 0.991 $\pm$ 0.001 \\
\hline
Total & 0.0552 $\pm$ 0.0005 \\
\hline
\end{tabular}
\end{center}
\label{tab:EffSum}
\end{table}

%\newpage
The ratio $R_{\epsilon}$ in Eq.~(\ref{eq:BRATIO}) accounts for
several effects all depending on the global properties of the event: 
trigger, on-line filter, event classification, T$_0$ determination, 
$K_L$-crash and $K_S$ identification. 
The various contributions to $R_\epsilon$ evaluated with simulation 
are listed in Table~\ref{tab:EffRatio} where only the statistical errors are shown. 
\begin{table}[htp]
\caption{Contributions to the ratio of efficiencies $R_{\epsilon}$ in 
Eq.~(\ref{eq:BRATIO}).
The error on $R_{\epsilon}$ is calculated as
the quadratic sum of the errors of the single ratios.}
\label{tab:EffRatio}
\begin{center}
\begin{tabular}{lccc}
Selection & $R_{\epsilon}$ \\ 
\hline
Trigger & 1.0649 $\pm$ 0.0005  \\
On-line filter & 1.0113 $\pm$ 0.0002 \\
Event classification & 1.1406 $\pm$ 0.0007 \\
T$_0$ determination & 1.0135 $\pm$ 0.0002\\
$K_L$-crash and $\beta^*$ & 1.1283 $\pm$ 0.0022\\
$K_S$ identification & 1.0481 $\pm$ 0.0012\\
\hline
$R_{\epsilon}$ & 1.472 $\pm$ 0.004 \\
\hline
\end{tabular}
\end{center}
\end{table}

The efficiency of the $K_S \to \pi^+\pi^-$ normalisation sample 
is measured using the preselected data 
by varying the cut on the vertex transverse position, as in Eq.~(\ref{eq:Vertex}), 
in 1 cm steps from $\rho^{\rm max}_{\rm vtx} = 1$ cm to 
$\rho^{\rm max}_{\rm vtx} = 4$ cm,
based on the observation that $\rho_{\rm vtx}$ and the tracks momenta are
very loosely correlated. Using Eq.~(\ref{eq:EFFICIENCY}) and extrapolating to 
$\rho^{\rm max}_{\rm vtx} = 5$ cm the efficiency is 
$\epsilon_{\pi\pi} = (96.569 \pm 0.004) \%$. 
Alternatively, the efficiency is evaluated using the $K_S \to \pi^+\pi^-$ data sample
(with $\rho^{\rm max}_{\rm vtx} = 5$ cm ):
$\epsilon_{\pi\pi}  = (96.657 \pm 0.002)\%$.
The latter value is used as the efficiency and the difference between the two values is taken as systematic uncertainty.  
The number of 
$K_S \to \pi^+ \pi^-$ events corrected for the efficiency is
\[ N_{\pi\pi}/\epsilon_{\pi\pi} = (292.08 \pm 0.27)\times10^6 . \]

\section{Systematic uncertainties} \label{SYSTEMATICS}

Three main systematic uncertainties affect the signal count: 
BDT and time-of-flight selection, and the $m^2_{\mu}$ fit. 

The distributions of the BDT classifier
output for the data and simulated signal and control sample events are in good 
agreement as shown in Figures~\ref{fig:BDToutput} 
and~\ref{fig:BDTLS-MC}.
The resolution of the BDT variable predicted by simulation comparing
the reconstructed events with those at generation level is
$\sigma_{\rm BDT} = 0.005$. The analysis is repeated varying the BDT cut of 
Eq.~(\ref{eq:BDTcut}), the number of signal events is found to be stable and
the rms value of the differences corresponds to a relative uncertainty of 0.3\%.

The main source of uncertainty in the TOF selection is the
cut on $\delta t_{\pi\pi}$ in Eq.~(\ref{eq:deltaTOFpipicut})
because the signal and background distributions in 
Figure~\ref{fig:dTOFpipi}(left) are steep and with opposite slopes;
the subsequent selection on $\delta t_{\mu}$ in Eq.~(\ref{eq:deltaTOFmu})
has a minor effect.
The resolution of the $\delta t_{\pi\pi}$ variable evaluated with simulation 
and $K_S \to \pi^+\pi^-$ data control samples is $0.27$ ns. 
The analysis is repeated varying the $\delta t_{\pi\pi}$ lower cut in the range 0.5--1.25 ns, 
the rms value of the differences
corresponds to a relative uncertainty of 3.0\%. 
This is the main systematic uncertainty affecting the measurement.

The fit to the $m_{\mu^2}$ distribution in Figure~\ref{fig:m2FIT} is repeated varying the range and the bin size. The relative systematic uncertainty on the number of events is 0.3\% obtained shifting the range by five bins on either side and increasing the bin size by a factor of two.

The statistical uncertainties of the simulation signal and control samples, and of the data control sample contribute with a  relative uncertainty of 0.8\% (Table~\ref{tab:EffSum}).
% Inserito in Tabella 4. The statistical uncertainties of the simulation signal and control samples contribute with a relative uncertainty of 0.8\%. 

The dependence of $R_{\epsilon}$ on systematic effects has been studied 
in previous analyses for different $K_S$ decays selected with the $K_L$-crash method: 
$K_S \to \pi^+ \pi^-$ and $K_S \to \pi^0 \pi^0$~\cite{ref:KStopp},
and $K_S \to \pi e \nu$~\cite{ref:KStopienu}.
The systematic uncertainties are evaluated by a comparison of data 
with simulation, as described in the following.

{\bf Trigger --} Two triggers are used for recording the events, 
the calorimeter trigger and the drift chamber trigger. The validation of the 
MC relative efficiency is derived from the comparison of the single-trigger
and coincidence rates with the data. The data over MC ratio is 0.999
with negligible error. 

{\bf On-line filter} and {\bf event classification --} 
The event classification produces
different streams for the analyses. The $K_L K_S$ stream selects 
events based on the properties of $K_S$ and $K_L$ decays.
In more than 99\% of the cases the events are selected based on the
$K_S$ decay topology and the $K_L$-crash signature and differences
between MC and data are accounted for in the systematic uncertainties 
derived below for the $K_S$ identification and $K_L$-crash.

{\bf T$_0$ --} 
%The trigger time is synchronised with the r.f. signal and the event T$_0$ is re-defined after event reconstruction.
The systematic uncertainty is evaluated analising the 
data and MC  T$_0$ distributions for the decays with the
most different timing properties: 
$K_S \to \pi^+\pi^-$ and $K_S \to \pi^0\pi^0$~\cite{ref:KStopp}.
The data over MC ratios is one with an uncertainty of less than 0.1\%.

{\bf $K_L$-crash} and {\bf $\beta^*$ selection --} The systematic uncertainty
is evaluated comparing data and simulated events tagged by 
$K_S \to \pi^+\pi^-$ and $K_S \to \pi^0\pi^0$ decays which have
different timing and topology characteristics. 
The data over MC ratio is 1.001 with negligible error.

{\bf $K_S$ identification --}  The systematic uncertainty due to the requirement 
of two tracks forming a vertex in the cylinder defined by Eq.~(\ref{eq:Vertex})  
is evaluated separately for signal and normalisation samples. The first is evaluated with $K_L \to \pi \mu \nu$ events 
selected with the same vertex requirements as for the signal but tagged by $K_S \to \pi^0\pi^0$ decays. 
For the $K_S \to \pi^+\pi^-$ sample the efficiency is evaluated by tagging with $K_L$-crash and removing the requirement of the vertex.
Combining the two values gives a data over MC ratio of
1.002 $\pm$ 0.017 where the error is due to the purity of the samples.    

The $R_{\epsilon}$ total systematic uncertainty is estimated by combining 
%the differences from one of 
the data over MC ratios and amounts to 1.7\%.

Including the systematic uncertainties the factors in Eq.~(\ref{eq:BRATIO}) are:
%\begin{equation}
\[ \epsilon_{\pi\mu\nu} = 0.0552 \pm 0.0017 
\quad {\rm and} \quad
R_{\epsilon} = 1.472 \pm 0.025. \]
%\label{eq:EffFinal}
%\end{equation}
All systematic uncertainties are summarised in Table~\ref{tab:Syst}.
\begin{table}[htp!]
\caption{Summary of systematic uncertainties of 
$\epsilon_{\pi\pi}$, $\epsilon_{\pi\mu\nu}$ and $R_{\epsilon}$.}
\begin{center}
\begin{tabular}{lccc}
Source & $\epsilon_{\pi\pi}$ [ \% ] & $\epsilon_{\pi\mu\nu}$ [ \% ] & $R_{\epsilon}$ [ \% ] \\
\hline
$K_S \to \pi^+\pi^-$ selection & 0.1 & & \\
\hline
BDT selection		& &	0.3 & \\
TOF selection		& &	3.0 & \\
Fit $m^2_{\mu}$ distribution &	& 0.3 & \\
MC and data CS statistics & & 0.8 &\\
%MC samples statistics & &  0.8 & \\
\hline
Trigger			&  & & \ \ 0.1 \\
T$_0$	determination	 & & &  $<$0.1 \\
$K_L$-crash and $\beta^*$	& &  &	\ \ 0.1 \\
$K_S$ identification	 & & & \ \ 1.7 \\
\hline
Total 		& 0.1	&  3.1  &  \ \ 1.7  \\
\hline
\end{tabular}
\end{center}
\label{tab:Syst}
\end{table}

\section{Result} \label{RESULT}
From Eq.~(\ref{eq:BRATIO}) with $N_{\pi\mu\nu} = 7223 \pm 180$,
$N_{\pi\pi}/\epsilon_{\pi\pi} = (292.08 \pm 0.27) \times 10^6$,
the values of the efficiencies 
%in Eq.~(\ref{eq:EffFinal}), 
$\epsilon_{\pi\mu\nu} = 0.0552 \pm 0.0017$,
$R_{\epsilon} = 1.472 \pm  0.025$,
and the value 
$\mathcal{B}(K_S \to \pi^+ \pi^-) = 0.69196 \pm 0.00051$ 
measured by KLOE~\cite{ref:KStopienu}, we derive the branching 
fraction
\[ \mathcal{B}(K_S \to \pi \mu \nu) = (4.56 \pm 0.11_{\rm stat} \pm 0.17_{\rm syst} )\times10^{-4}
= (4.56 \pm 0.20)\times10^{-4} . \]
%
%This is the first measurement of this decay mode. 
%Assuming universality of the kaon--lepton coupling, 
%the expected value~\cite{ref:PDG} is 
%\[ \mathcal{B}(K_S \to \pi \mu \nu) = \mathcal{B}(K_S \to \pi e \nu) \times %R(I^{\ell}_K) 
%\times \frac{1 + \delta^{\pi\mu\nu}_{\rm LD}}{1 + \delta^{\pi e \nu}_{\rm LD}} %
%= (4.69 \pm 0.06)\times10^{-4}  \]
%as derived from the value of the branching fraction $\mathcal{B}(K_S \to \pi e \nu)$
%%$\mathcal{B}(K_S \to \pi e \nu) = (7.046 \pm 0.091)\times10^{-4}$
%measured by KLOE~\cite{ref:KStopp}, the ratio $R(I^{\ell}_K)$
%%$R(I^{\ell}_K) = 0.6622 \pm 0.0018$ 
%of the phase space 
%integrals for the semileptonic decays $K_L \to \pi \mu \nu$ and $K_L \to \pi e %\nu$ measured by KTeV~\cite{ref:KTeV}, and the contributions of long-distance radiative correction semileptonic kaon decays $\delta_{LD}$.~\cite{ref:Andre}.
%
%\noindent
This is the first measurement of this decay mode and completes the set of
kaon semileptonic decays. The branching fraction for $K_S (K_L) \to \pi \ell \nu$ decay is related to the weak coupling constant and $V_{us}$ 
through the relation 
\begin{equation}
\label{eq:formula}
\mathcal{B}(K_S \to \pi \ell \nu) = 
\frac{G^2 |f_+(0) V_{us}|^2}{192 \pi^3}\ \tau_s m_K^5 I_{K}^{\ell} S_{\rm EW} (1 + \delta_{\rm EM}^{K\ell})^2
\end{equation}
where $f_+(0)$ is the hadronic form factor at zero momentum transfer,
$m_K$ and $\tau_S$ are the $K_S$ mass and lifetime, $I_{K}^{\ell}$ is the phase space integral, $S_{\rm EW}$ 
is the short-distance electroweak correction \cite{ref:MarcianoSirlin} and $\delta^{K\ell}_{\rm EM}$ is the long-distance electromagnetic correction~\cite{ref:Cirigliano2008, ref:Andre}.
Assuming universality of the kaon--lepton coupling the expected value~\cite{ref:PDG} is 
\[ \mathcal{B}(K_S \to \pi \mu \nu)_{\rm PDG} = \mathcal{B}(K_S \to \pi e \nu) \times R(I^{\ell}_K) 
 \times \frac{(1 + \delta^{K\mu}_{\rm EM})^2}{(1 + \delta^{K e}_{\rm EM})^2} 
= (4.69 \pm 0.06)\times10^{-4} \]
as derived from the value of the branching fraction $\mathcal{B}(K_S \to \pi e \nu)$  
measured by KLOE~\cite{ref:KStopienu} and the ratio $R(I^{\ell}_K)$ 
of the phase space integrals for the semileptonic decays 
$K_L \to \pi \mu \nu$ and $K_L \to \pi e \nu$ measured by KTeV~\cite{ref:KTeV}. 

Inverting Eq.~(\ref{eq:formula}) and using $I_K^\mu = 0.10262 \pm 47$~\cite{ref:KLOE|Vus|} we derive 
\[ |f_+(0) V_{us}|_{K_S\to\pi \mu \nu} = 0.2126 \pm 0.0046 . \] 
The ratio to the value for $K_S \to \pi e \nu$ decay, $|f_+(0) V_{us}|_{K_S \to \pi e \nu} = 0.2153 \pm 14$~\cite{ref:KLOE|Vus|}, allows to confirm the assumption of kaon--lepton coupling universality
\[  r_{\mu e} =
 \frac{|f_+(0) V_{us}|^2_{K_S\to\pi \mu \nu}}{|f_+(0) V_{us}|^2_{K_S\to\pi e \nu}} = 0.975 \pm 0.044 . \]
These results are consistent with those determined for the other kaon semileptonic decays \cite{ref:Antonelli2010, ref:KLOE|Vus|} though less precise mainly due to the intrinsic limitations related to $\mu$--$\pi$ discrimination in the momentum range 100--250 MeV.

\section{Conclusion} \label{CONCLUSION}
A measurement of the 
branching fraction for the decay $K_S \to \pi \mu \nu$ is presented
based on data collected with the KLOE experiment at the DA$\Phi$NE $e^+e^-$ 
collider corresponding to an integrated luminosity of 1.6 fb$^{-1}$.
The $\phi \to K_LK_S$ decays are exploited to select samples of pure and 
quasi-monochromatic
$K_S$ mesons and data control samples of $K_L \to \pi \mu \nu$ decays. 
The $K_S \to \pi \mu \nu$ decays are selected by
a boosted decision tree built with kinematic variables and by a measurement of time-of-flight.
The efficiencies for detecting the $K_S \to \pi \mu \nu$ decays are derived from 
$K_L\to \pi \mu \nu$ data control samples. 
A fit to the $m_{\mu}^2$ distribution finds $7223 \pm 180$ signal
events. Normalising to $K_S \to \pi^+ \pi^-$ decay events,   
the result for the branching fraction is 
$\mathcal{B}(K_S \to \pi \mu \nu) = (4.56 \pm 0.11_{\rm stat} \pm 0.17_{\rm syst})\times10^{-4}$
to be compared with the expected value of 
$(4.69 \pm 0.06 )\times10^{-4}$ assuming lepton-flavour universality.

\vskip 1.cm
\small{
{\bf Acknowledgements -- }
We warmly thank our former KLOE colleagues for the access to the data collected
during the KLOE data taking campaign.
We thank the DA$\Phi$NE team for their efforts in maintaining low background 
running conditions and their collaboration during all data taking. 
We thank our technical staff: 
G.F. Fortugno and F. Sborzacchi for their dedication in ensuring efficient operation 
of the KLOE computing facilities; 
M. Anelli for his continuous attention to the gas system and detector safety; 
A. Balla, M. Gatta, G. Corradi and G. Papalino for electronics maintenance; 
C. Piscitelli for his help during major maintenance periods. 
This work was supported in part 
by the Polish National Science Centre through the Grants No.
2013/11/B/ST2/04245, 
2014/14/E/ST2/00262, 
2014/12/S/ST2/00459, 
2016/21/N/ST2/01727, 
2016/23/N/ST2/01293, 
2017/26/M/ST2/00697.

}

\newpage
%\appendix{\bf Appendix}

%\begin{table}[htp!]
%\caption{Summary of systematic uncertainties of 
%$\epsilon_{\pi\pi}$, $\epsilon_{\pi\mu\nu}$ and %$R_{\epsilon}$.}
%%\begin{center}
%\begin{tabular}{lccc}
%Source & $\epsilon_{\pi\pi}$ [ \% ] & %$\epsilon_{\pi\mu\nu}$ [ \% ] & $R_{\epsilon}$ [ \% ] \\
%\hline
%$K_S \to \pi^+\pi^-$ selection & 0.1 & & \\
%\hline
%BDT selection		& &	0.3 & \\
%TOF selection		& &	3.0 & \\
%Fit $m^2_{\mu}$ distribution &	& 0.3 & \\
%MC samples statistics & &  0.8 & \\
%\hline
%Trigger			&  & & 0.1 \\
%T$_0$	determination	 & & &  $<$0.1 \\
%$K_L$-crash and $\beta^*$	& &  &	0.1 \\
%$K_S$ identification	 & & & 1.7 \\
%\hline
%Total 		& 0.1	&  3.1  &  1.7  \\
%\hline
%\end{tabular}
%\end{center}
%\label{}
%\end{table}

\end{document}